\documentclass[%
 reprint,
 amsmath,amssymb,
 aps,
 prl
]{revtex4-2}

\usepackage{graphicx}
\usepackage{dcolumn}
\usepackage{bm}
\usepackage{xcolor}
\usepackage{braket}
\usepackage{lipsum}
\usepackage{hyperref}

\begin{document}

\preprint{APS/123-QED}

\title{Light-Hole Spin Qubits in Strained SiGe Lattice-Matched to Ge
}

\author{Edmondo Valvo}
\email{e.valvo@tudelft.nl}
\affiliation{QuTech and Kavli Institute of Nanoscience, Delft University of Technology, Delft, The Netherlands}

\author{Davide Costa}
\affiliation{QuTech and Kavli Institute of Nanoscience, Delft University of Technology, Delft, The Netherlands}
\email{d.costa@tudelft.nl}

\author{Patrick Del Vecchio}
\affiliation{QuTech and Kavli Institute of Nanoscience, Delft University of Technology, Delft, The Netherlands}
\email{p.delvecchio@tudelft.nl}

\author{Stefano Bosco}
\affiliation{QuTech and Kavli Institute of Nanoscience, Delft University of Technology, Delft, The Netherlands}
\email{s.bosco@tudelft.nl}
\author{Giordano Scappucci}
\affiliation{QuTech and Kavli Institute of Nanoscience, Delft University of Technology, Delft, The Netherlands}
\email{g.scappucci@tudelft.nl}
\author{Maximilian Rimbach-Russ}
\affiliation{QuTech and Kavli Institute of Nanoscience, Delft University of Technology, Delft, The Netherlands}
\email{m.f.russ@tudelft.nl}

\date{\today}

\begin{abstract}
Strained germanium ($\varepsilon$-Ge) quantum wells on metamorphic SiGe buffers have enabled advanced hole-based spin qubit devices. Alternatively, unstrained Ge with lattice-matched strained silicon-germanium ($\varepsilon$-SiGe)
barriers eliminates the need for metamorphic buffers altogether. The ground state character of both these platforms is predominantly heavy-hole (HH) with a largely anisotropic spin response. We propose and study an alternative heterostructure, lattice-matched to Ge, in which both the SiGe quantum well and barriers are tensile strained, with their composition contrast providing the band offset for confinement and the tensile strain stabilizing a light-hole (LH) ground state.
We show large spin-orbit coupling (SOC), both linear and cubic, along with a significantly more isotropic spin response compared to strained HH qubits. We also study the decoherence properties of the proposed device, showing an appreciable gain in the quality factor compared to their HH counterparts. Finally, we propose a bilayer heterostructure that allows for electrical switching between HH and LH ground state character.

\end{abstract}
\maketitle
\textit{Introduction---}Hole spin qubits in strained germanium ($\varepsilon$-Ge) are an established platform for quantum computing~\cite{kloeffelProspectsSpinBasedQuantum2013,watzingerGermaniumHoleSpin2018,hendrickxFourqubitGermaniumQuantum2021,fangRecentAdvancesHolespin2023,lawrieSimultaneousSinglequbitDriving2023,Hendrickx2024,johnTwodimensional10qubitArray2025a,saez-mollejoExchangeAnisotropiesMicrowavedriven2025,kellyIdentifyingMitigatingErrors2025,dijkemaSimultaneousOperation18qubit2026,vanriggelen-doelmanCoherentSpinQubit2024,ademiDistributingEntanglementDistant2025,tsoukalasDressedSinglettripletQubit2026,seidlerSpatialUniformityGtensor2025}, offering full electrical driving and the prospect of baseband control~\cite{Wang2024Hopping,rimbach-russGaplessSingleSpinQubit2025,zhangUniversalControlFour2025,boscoExchangeOnlySpinOrbitQubits2024a,nguyenDegenerateSingletTripletQubit2026}. The compressive strain imparted by the metamorphic SiGe buffer separates the Light-Hole (LH) subbands from the Heavy-Hole (HH) manifold, stabilizing the computational subspace at the cost of a strongly anisotropic spin response between growth and planar directions. That same metamorphic substrate, however, introduces threading and misfit dislocations and strain fluctuations that render the g-tensor non-uniform~\cite{corley-wiciakNanoscaleMapping3D2023,valvoElectricallyTuneableVariability2025,martinez2025variabilityholespinqubits,martinezDisorderindependentHoleSpin2026}. This has motivated buried Ge wells with strained SiGe barriers (Ge/$\varepsilon$-SiGe), lattice-matched to an underlying Ge substrate, which retain large SOC while reducing dot-to-dot variability~\cite{costaBuriedUnstrainedGermanium2025,mauroHoleSpinQubits2025}.

Tensile-strained Ge wells offer a complementary route in which the LH becomes the ground state~\cite{vecchioLightHoleGateDefinedSpinOrbit2023,vecchioLightholeSpinConfined2024,vecchioFullyTunableStrong2025,dsouzaHeavyHoleVs2025,abadillo-urielSpinQubitManipulation2017,gyorgyQuantumGeometricalDescription2026}. The leading candidate, $\varepsilon$-Ge/GeSn, exploits the larger lattice constant of a GeSn strain-relaxed buffer to raise the LH band, granting fast driving, large spin-orbit coupling (SOC), strong spin-photon coupling in the microwave regime, and a direct bandgap for optical photon interfaces~\cite{vecchioLightHoleGateDefinedSpinOrbit2023,abadillo-urielSpinQubitManipulation2017,moghaddamExportingSuperconductivityGap2014}. The reliance on a defective metamorphic buffer, however, persists.

\begin{figure}
    \centering
    \includegraphics[width=\linewidth]{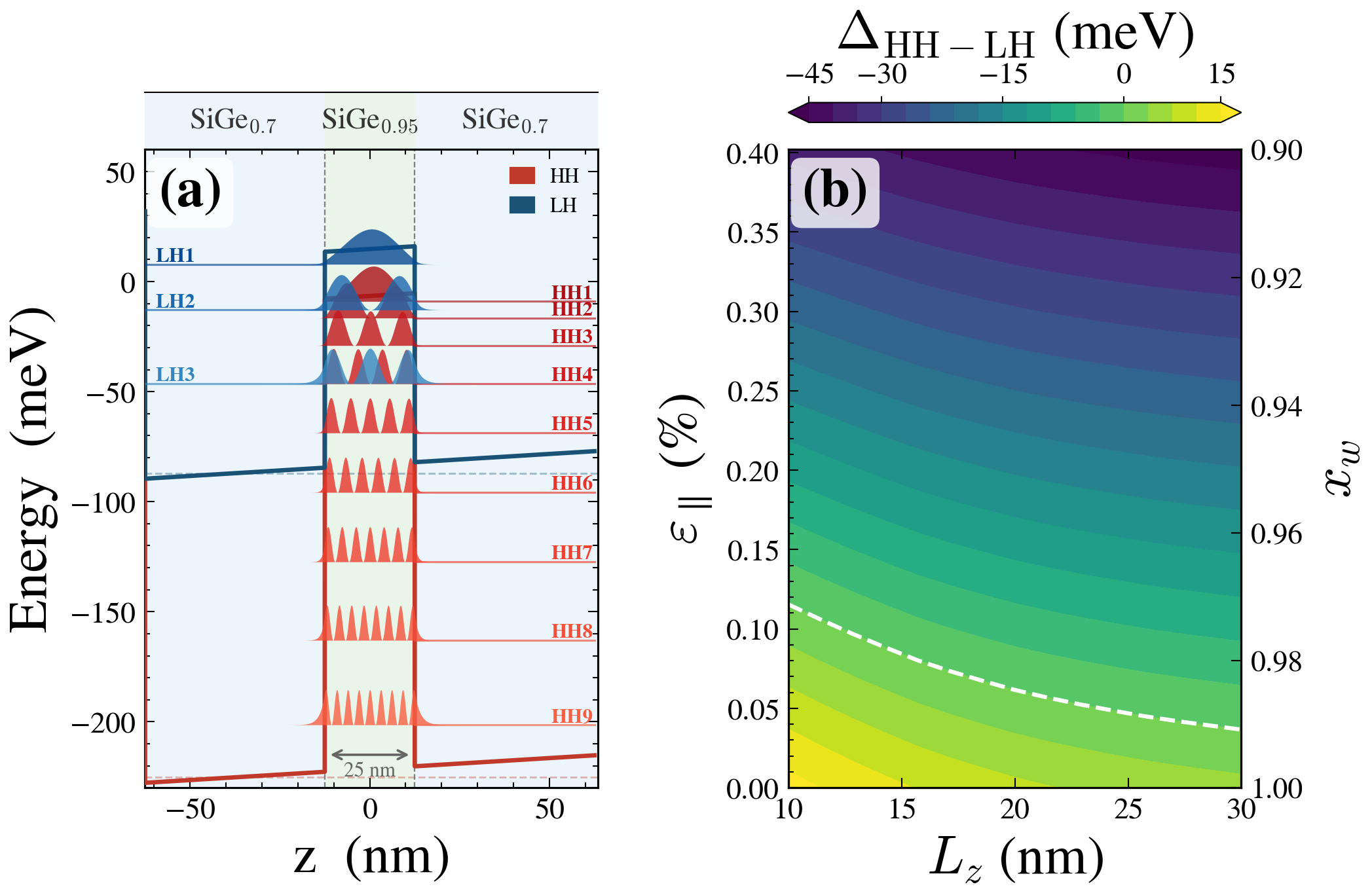}
    \caption{We study a heterostructure lattice matched to bulk Ge comprising stacked $\text{Si}_{1-x}\text{Ge}_x$ layers of different composition leading to a tensile-strained quantum well. We show in (a) an exemplary parameter choice with a relaxed Ge buffer at the left-most edge and then an alternation of $\text{Si}_{0.30}\text{Ge}_{0.70}$/$\text{Si}_{0.05}\text{Ge}_{0.95}$/$\text{Si}_{0.30}\text{Ge}_{0.70}$. In (b) a map of the energy gap between HH ground state and LH ground state as a function of well thickness and Ge concentration in the SiGe well. The dashed white line represents the transition line where the HH-LH gap closes. The strain $\varepsilon_\parallel$ is computed from the linear Vegard interpolation of the lattice constant of the $\text{Si}_{1-x_w}\text{Ge}_{x_w}$ well.}
    \label{fig:Delta}
\end{figure}

\begin{figure*}
    \centering
    \includegraphics[width=\textwidth]{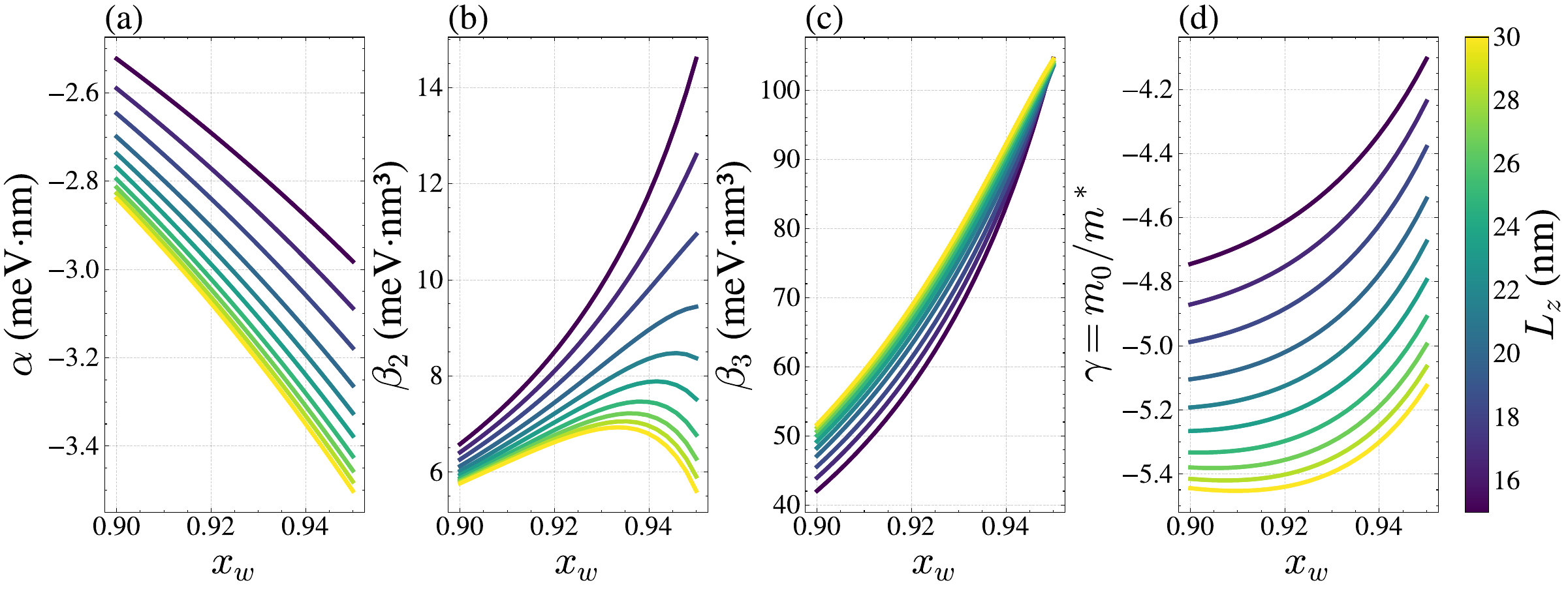}
    \caption{Subband parameters as a function of Germanium concentration ($x_w$) in the SiGe well and well thickness. (a) linear Rashba coefficient $\alpha$, (b) cubic Rashba coefficient $\beta_2$ proportional to $k_+^3$ in the folded Hamiltonian, (c) cubic Rashba coefficient $\beta_3$ proportional to $k_-k_+k_-$ in the folded Hamiltonian and finally the effective mass of the lowest subband in (d). All four figures were extracted for a vertical field of $1$ MV/m and barrier composition $\text{Si}_{0.3}\text{Ge}_{0.7}$}
    \label{fig:Rashba_and_Mass}
\end{figure*}

In this work we show that a tensile-strained LH ground state can be engineered with strained SiGe layers alone, by setting the lattice constant of the entire stack to that of germanium and without integrating other group IV materials such as Sn. The stack, grown on a Ge wafer, alternates Si$_{1-x}$Ge$_x$ alloys of different chemical composition to host the quantum well (Fig.~\ref{fig:Delta}(a)): The chemical contrast gives the band offset for confinement, while the strain profile stabilizes the lowest LH subband as the ground state. We find that the accumulated two-dimensional hole gas (2DHG) supports sizeable linear and cubic spin-orbit couplings and a dipole moment for microwave driving orders of magnitude larger than in conventional HH devices. The planar g-tensor is enhanced by about an order of magnitude, considerably reducing the anisotropy that characterizes HH-based devices ($g^{*,\text{HH}}_\parallel \approx 0.1$, $g^{*,\text{HH}}_\perp\approx 15$). Furthermore, the gain in dipole moment outweighs the reduced coherence, yielding an order-of-magnitude improvement in quality factor over typical $\varepsilon$-Ge qubits. Finally, to circumvent reduced coherence, combining a lower Ge well with a top tensile-strained SiGe well allows the ground state to be switched between HH and LH character by the vertical electric field alone. A complementary switchable HH--LH qubit, based on two Ge wells with SiGeSn barriers and gate-controlled shuttling, is proposed in concurrent work~\cite{GyorgySwitchableHHLH}. Taken together, these results position tensile-strained SiGe heterostructures as a practical strategy for the design of germanium hole spin qubits.\\ \textit{Microscopic Model---} We model the system via the methodology introduced in~\cite{vecchioTailoringGermaniumHeterostructures2026},  which first consists of a numerical diagonalization of the 6-band $k\cdot p$ Hamiltonian at zero magnetic field $\mathbf{B}$ and in-plane momentum $\mathbf{k}_\parallel$. We use the resulting eigenstates as the basis set to construct the Hamiltonian at finite $\mathbf{B} = B (\sin{\theta}\cos{\phi} \textbf{e}_x+ \sin{\theta}\sin{\phi} \textbf{e}_y+ \cos{\theta}\textbf{e}_z)$ and finite $\mathbf{k}_\parallel$. This yields the projected QW Hamiltonian
\begin{equation}\label{HQW}
\begin{aligned}
\mathcal{H}=&  
\mathbf{E}_0^{\mathrm{qw}}+ V_\parallel(x,y) \\
&+ \alpha_0\big[\mathbf{M}_\gamma K_\parallel^2 + \frac{\cos\theta}{2l_B^2}\mathbf{M}_g\\
&+ \frac{\sin\theta}{2l_B^2}\left(\frac{\sin\theta}{2l_B^2} \mathbf{N}_\gamma r_\phi  - \tilde{\mathbf{N}}_\gamma\right) r_\phi\big] \\
&+ \alpha_0[i\mathbf{M}_1 K_- + \mathbf{M}_2 K_-^2 \\ 
&+ \frac{\sin\theta}{2l_B^2}\left(e^{-i\phi}\mathbf{N}_g + i\mathbf{N}_1^+ r_\phi K_- + i\mathbf{N}_1^- K_- r_\phi\right) + \mathrm{h.c.}].
\end{aligned}
\end{equation}
 The diagonal matrix $\mathbf{E}_0^{\mathrm{qw}}$
collects the subband energies $E^{\mathrm{H}}$ and $E^{\eta}$ of the
quantum well at $\mathbf{K}_\parallel=0$ with the $H$ and $\eta$ labels representing the HH and LH/Split-Off (SO) bands respectively. Further details are specified in the Supplementary Material (SM)~\cite{supplemental_material}. Moreover $\textbf{K}=k + \frac{e}{\hbar}A$, $\textbf{K}_\pm = \textbf{K}_x \pm i \textbf{K}_y$ and $l_B=\sqrt{\hbar/(eB)}$, where $B = \sqrt{\textbf{B}\cdot \textbf{B}}$. The prefactor
$\alpha_0=\hbar^2/2m_0$ sets the overall energy scale, and all matrices
$\mathbf{M}$, $\mathbf{N}$ are overlap integrals of the
growth-direction envelopes $f_j^{l,s,h}$ with $f^l$, $f^s$ and  $f^h$ the Light-Hole, Split-Off, and Heavy-Hole envelopes of the $j$-th spinor. Here the information pertaining to strain, band-offset, and out-of-plane electric field, $eF_z$, is encoded in the envelopes and eigenenergies of the subbands, which is why these terms do not explicitly appear in Eq.~\ref{HQW}.
\begin{figure}
    \centering
    \includegraphics[width=\linewidth]{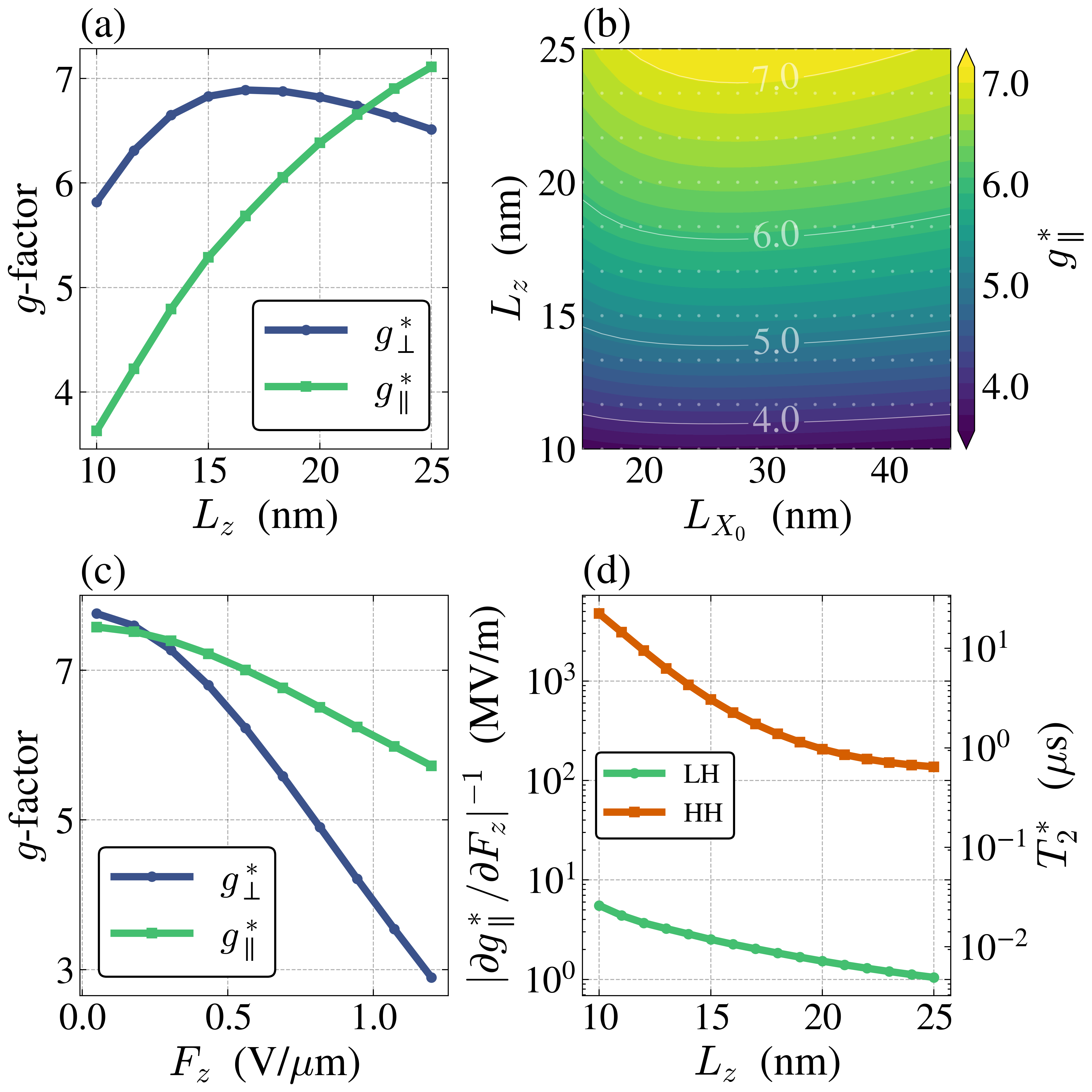}
    \caption{Simulated g-factors in the growth direction ($g^*_{\perp}$) and purely in plane along the $x$ direction ($g^*_{\parallel}$) at an electric field $F_z = 0.5$ MV/m for (a) a sweep of the $\text{Si}_{0.05}\text{Ge}_{95}$ well thickness ($L_z$) with fixed planar confinement at 35 nm, (b) isotropic planar confinement $L_{X_0}$ and well thickness  and (c) out-of-plane linear electric field $F_z$ for $L_z=25$ nm and $L_{X_0}=35$ nm.  The barrier composition is set to $\text{Si}_{0.30}\text{Ge}_{0.70}$. (d) The inverse of the derivative of the in-plane g-factor with respect to the vertical electric field (and corresponding dephasing time) is computed as a function of well thickness for the proposed tensile strained heterostructure with $\text{Si}_{0.05}\text{Ge}_{95}$ well and a typical $\varepsilon$-Ge device with stack $\text{Si}_{0.20}\text{Ge}_{80}$/$\text{Ge}$/$\text{Si}_{0.20}\text{Ge}_{80}$ and relaxed barriers.}
    \label{fig:g}
\end{figure}
The first bracket contains the terms diagonal in the orbital angular
momentum channel. $\mathbf{M}_\gamma K_\parallel^{2}$ is the isotropic
in-plane kinetic energy. The
term proportional to $\cos\theta\,\mathbf{M}_g/2l_B^{2}$ is linear in the
out-of-plane field component $B\cos\theta$ and encodes the Zeeman response. The remaining contribution is generated by the
in-plane component $B\sin\theta$ through the gauge choice
$\textbf{A}= \frac{B\cos{\theta}}{2}(-y \textbf{e}_x + x\textbf{e}_y)-Br_\phi\sin\theta \textbf{e}_z$, with $r_\phi=x\sin\phi-y\cos\phi$.
The second bracket collects the off-diagonal couplings, which mix the H
and $\eta$ blocks and are the source of the anisotropic and
spin-dependent physics, comprising $i\mathbf{M}_1K_-$ and $\mathbf{M}_2K_-^{2}$. The terms proportional to $\sin\theta/2l_B^{2}$ mix the in-plane field
with orbital motion: $e^{-i\phi}\mathbf{N}_g$ is the planar Zeeman term, while
$i\mathbf{N}_1^{\pm}$ are the mixed field-momentum terms. Finally, $V_\parallel(x,y)$ is the
electrostatic in-plane confinement, set to a harmonic potential for the present work. 
Further details on the microscopic model are presented in the SM and follows closely previous works~\cite{vecchioTailoringGermaniumHeterostructures2026}.

\textit{Tensile Strained SiGe wells---}
The most straightforward way to obtain a tensile strained SiGe well is to grow a strained-SiGe heterostructure lattice-matched to germanium. The heterostructure comprises  $\text{Si}_{1-x_b}\text{Ge}_{x_b}/ \text{Si}_{1-x_w}\text{Ge}_{x_w}$ with $x_b$ the barrier's Ge stoichiometry and $x_w$ that of the well. In the 2D limit (assuming no planar confinement) Fig.~\ref{fig:Delta}b) shows the energy gap between the first Heavy-Hole and Light-Hole subbands ($\Delta_{\text{HH}-\text{LH}}$) varying $x_w$ and the thickness of the tensile strained well $L_z$, hence negative values of the gap correspond to a LH ground state. We observe for values $x_w< 0.97$ a Light-Hole ground state throughout the considered range of well thicknesses. A typical band alignment and well structure is presented in Fig.~\ref{fig:Delta}a), showing a well confined ground state and up to two excited states below the  Light-Hole continuum limit.

\textit{2DHG Parameters---} We now study the dependence of the subband parameters on the stoichiometry and composition of the heterostructure. In particular we first look at the band parabolicity $\gamma = m_0/m^*$, with $m^*$ the effective mass, and the Rashba spin-orbit coefficients $\alpha,\beta_2, \beta_3$ that quantify the magnitude of the linear and cubic spin-orbit coupling computed from the projection of the z-subband matrices of Eq.~\ref{HQW} on the lowest doublet~\cite{vecchioTailoringGermaniumHeterostructures2026}
\begin{equation}
    \tilde{\mathcal{H}} = \alpha_0 \gamma k^2_\parallel +  [i( \alpha k_- - \beta_2 k^3_+ - \beta_3 k_- k_+ k_- )\sigma_- + \text{h.c.}].
\end{equation}

In Fig.~\ref{fig:Rashba_and_Mass}a) we observe the dependence of the linear Rashba coefficient $\alpha$ on the Ge concentration in the well and the well thickness. We notice that the magnitude of this term follows the increase in Ge concentration in the well which boosts the value of the Luttinger parameter $\gamma_3$ (See the SM~\cite{supplemental_material}). At the same time the linear spin-orbit is increased by making the well wider. This can again be understood qualitatively by noting that another driver of linear spin-orbit coupling is the transition matrix element $\bra{f_j^l}k_z\ket{f_j^s}$  (See the SM~\cite{supplemental_material}).

In Fig.~\ref{fig:Rashba_and_Mass}b-c) we plot the cubic Rashba terms $\beta_2, \beta_3$ for the same parameter ranges. We observe a divergence for thinner wells when increasing the Ge concentration caused by the closing of the HH-LH gap. Importantly, in contrast with HH ground states, the $\beta_3$ parameter of LH states becomes proportional to the sum of the Luttinger parameters ($\gamma_2+\gamma_3$) and is thus larger than $\beta_2 \propto (\gamma_2-\gamma_3)$~\cite{delvecchioDynamicsSpinOrbitCoupling2024}.  
Finally in Fig.~\ref{fig:Rashba_and_Mass}d) we can observe the behaviour of the extracted effective mass for the lowest subband showing a smooth increase when closing the HH-LH gap. Moreover when increasing the well thickness the absolute value of the effective mass decreases moving towards the bulk limit.
Combining these results it becomes apparent that confining a hole spin in the proposed manner grants a significant increase in the linear and cubic $\beta_3$ SOC relevant for microwave driving with respect to typical planar HH ground states, consistent with results reported for LH states in GeSn~\cite{vecchioLightholeSpinConfined2024,vecchioLightholeSpinConfined2024}.

\textit{A Light-Hole qubit---}
Following the previous discussions we now explore the characteristics of a qubit hosted in the lowest Light-Hole doublet of the heterostructure outlined previously. To perform this we reintroduce the planar confinement set aside above, $V_\parallel=\alpha_0(\frac{x^2}{l_x^4}+\frac{y^2}{l_y^4})$, with $l_x$ and $l_y$ being the harmonic confinement lengths in the planar directions that we set equal for the remainder of this work $l_x = l_y = L_{X_0}$. As a first parameter we model the g-tensor which is obtained by diagonalizing Eq.~\ref{HQW} for $\textbf{B}=0$ and using the obtained eigenstates to project the linear terms in magnetic field arising from the Zeeman or orbital contributions~\cite{vecchioTailoringGermaniumHeterostructures2026}. This is in line with the g-tensor formalism~\cite{Venitucci2018}.

In Fig.~\ref{fig:g} we fix the well composition at $\text{Si}_{0.05}\text{Ge}_{95}$ and present the behavior of the in-plane and out-of-plane g-factors $g_{\perp,\parallel}^* = \frac{||g \textbf{B}_{\perp,\parallel}||}{||\textbf{B}_{\perp,\parallel}||}$ for a sweep of the relevant device parameters: well thickness ($L_z$), isotropic planar confinement $L_{X_0}$ and out-of-plane linear electric field $F_z$. In bulk LH states the behaviour of in-plane and out-of-plane g-factors is inverted with respect to HH, $g^*_{\perp}$ is smaller than $g^*_{\parallel}$~\cite{vecchioLightHoleGateDefinedSpinOrbit2023}. We observe this inversion for $L_z>22$ nm along with an interesting non-monotonic trend of the $g^*_{\perp}$ component with respect to the well width. This behaviour is mainly due to the HH-LH gap reaching the minimum set by strain while the $k_z$ transition matrix element between LH and HH states keeps decreasing, further details are reported in the SM~\cite{supplemental_material}.

In Fig.~\ref{fig:g} (b) we study the dependency of the planar g-factor $g_\parallel^*$ on the planar confinement $L_{X_0}$ and SiGe well thickness $L_z$. We note a flattening of the trends beyond $L_{X_0}=20$ nm indicating the possibility of sweet spots. In Fig.~\ref{fig:g} (c) we plot these quantities as a function of the out-of-plane electric field. We observe a  renormalization for both g-tensor components. In particular, $g^*_{\perp}$ drops from $\approx7$ to less than $3$. This trend can be largely attributed to the strong intra-band mixing of the LH/SO manifold. In the QW Hamiltonian the intra-band coupling of the LH/SO manifold is denoted as $T^\eta$. Increasing the electric field boosts this term which modifies the ground state composition. The crucial effect that manifests in LH ground states is the strong increase in $T^\eta$ which makes the ground state a mixed $J_z$ projection of $J_z = \pm1/2$ component at high electric fields. Moreover as a further renormalization, the orbital magnetic field correction give a significant variation as a function of electric field, further details are presented in the SM~\cite{supplemental_material}. This implies a rather strong tunability of the spin response with the imposed electric field. Finally in  Fig.~\ref{fig:g}d) we compare the  derivative of the in-plane g-factor with respect to the vertical electric field as a function of well thickness for the proposed tensile strained heterostructure with $\text{Si}_{0.05}\text{Ge}_{95}$ well and a typical $\varepsilon$-Ge HH  device with stack $\text{Si}_{0.20}\text{Ge}_{0.80}$/$\text{Ge}$/$\text{Si}_{0.20}\text{Ge}_{0.80}$ and relaxed barriers. As a secondary y-axis we show the corresponding dephasing time computed, in first approximation, by attributing all charge noise in the device to fluctuations in the out-of-plane electric field. Under this assumption we can write the dephasing time as~\cite{Wang2024Modelling} $T_2^* = \hbar\left[\mu_B \sqrt{\text{log}(r)}A_z |\frac{\partial g}{\partial F_z}B|\right]^{-1}$ with $r=1.68 \times 10^9$ being the ratio of upper and lower frequency cutoffs, $A_z = 10$ kV/m the amplitude of the out-of-plane field fluctuations, chosen to yield $T_2^* \approx 1 \mu s$ for a planar magnetic field in the $\varepsilon$-Ge hole spin qubit.

\begin{figure}
    \centering
    \includegraphics[width= \linewidth]{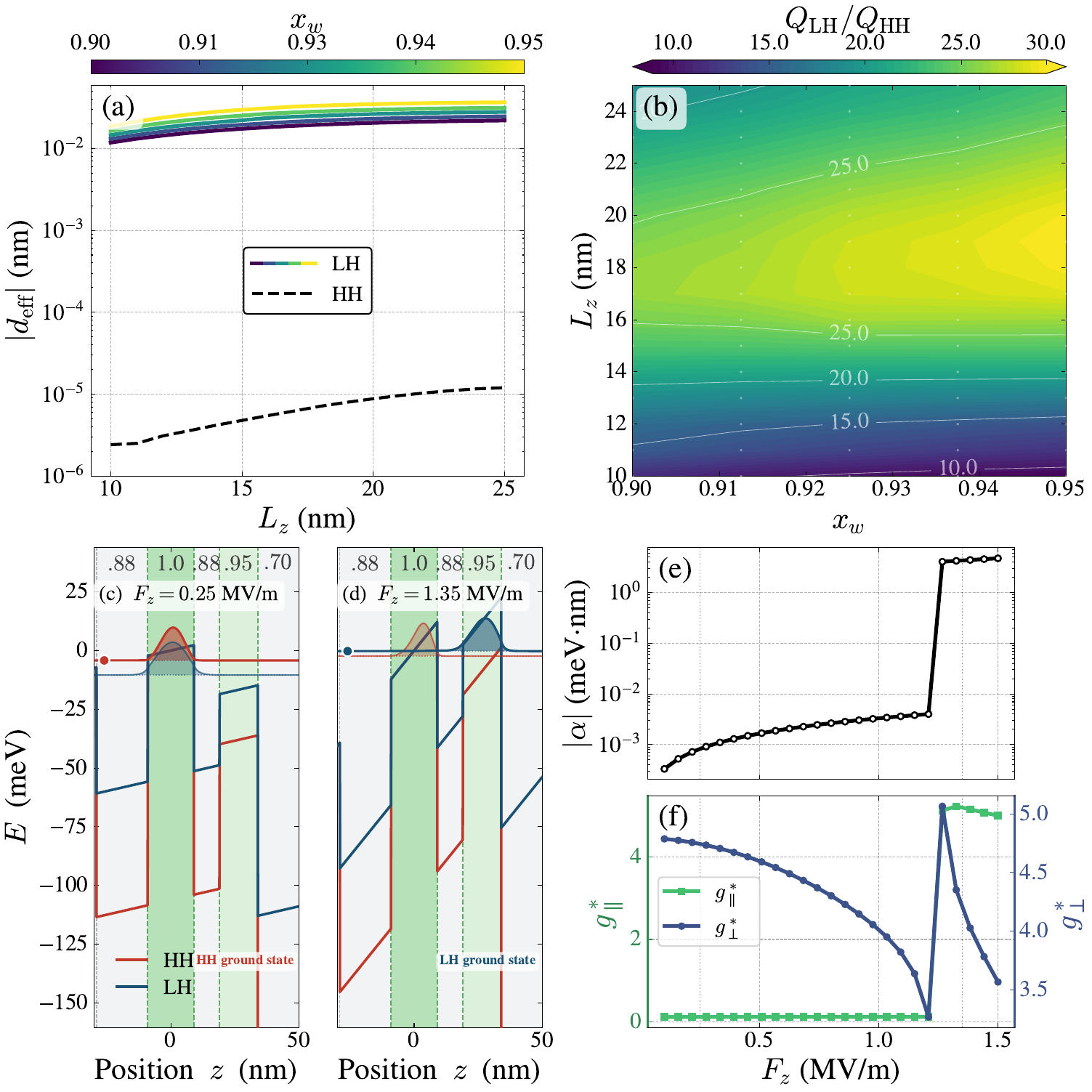}
    \caption{(a) Computed dipole moment as the transition matrix element of the position operator between the two computational states $d=\bra{0}x\ket{1}$, expressed in nm, for a planar confinement $L_{X_0} = 35$ nm, barrier composition $\text{Si}_{70}\text{Ge}_{30}$, vertical electric field $F_z = 1.2$ MV/m and magnetic field amplitude $15$ mT. Overlayed in black is the dipole moment of the $\varepsilon$-Ge HH heterostructure utilized for comparison. (b) Ratio of Quality factors $Q=T_2^* d_\text{eff}e E_{\text{AC}}/\hbar$ for a Light-Hole ground state in the tensile strained SiGe well ($Q_{LH}$) and a $\varepsilon$-Ge HH hole spin qubit ($Q_{HH}$) computed for fluctuations around a vertical field of $1.2$ MV/m. (c)-(d) Proposed heterostructure to achieve a HH-LH bilayer for $0.25$ MV/m and $1.35$ MV/m respectively, showing the transition from HH to LH ground state. The grey insets at the top of each layer represent the Ge concentration in each SiGe alloy, with the underlying substrate not shown but kept fixed at pure Ge like the monolayer study.  (e) Linear SOC parameter extracted as a function of out-of-plane electric field. Around $1.25$ MV/m we observe the sudden transition with a large boost in the strength of $\alpha$. Finally (f) shows the evolution of the g-factor for the same parameter range with the typical anisotropy inversion expected between HH and LH ground states.}
    \label{fig:deff}
\end{figure}

As the next step in our analysis we look to the dipole moment, defined as the position operator transition matrix element $d_\text{eff} = \bra{\uparrow}x\ket{\downarrow}$ with $\ket{\uparrow}$ and $\ket{\downarrow}$ being the Kramer pair states and is proportional to the Rabi frequency for weak electrical driving or spin-photon coupling. In Fig.~\ref{fig:deff} we plot the dipole moment for different Ge composition and thicknesses of the well. Importantly, we note a boost of three orders of magnitude with respect to the typical $\varepsilon$-Ge HH heterostructure. Furthermore we notice increasing the well thickness or decreasing the Si concentration in the well boost the dipole moment, due to increases interband couplings and reduced HH-LH gap.
\\
To compare different heterostructures, we define the quality factor measure $Q=T_2^*\times \text{Rabi}\equiv T_2^* d_\text{eff} e E_{AC}/\hbar$, with $E_{AC}$ being the same drive amplitude for both platforms. In Fig.~\ref{fig:deff}(b) we note at least an order of magnitude improvement of our proposed LH qubit with respect to the conventional $\varepsilon$-Ge hole qubit. We also see a large increase for wells around $18$ nm and larger amount of Ge in the SiGe well, while a degradation is observed for more well separated HH-LH states. This can naturally be understood by noting that the dipole moment increase in those regions starts to be marginal compared to the decay of the coherence times.

\textit{Switching of Ground State Composition in a Bilayer---} Even though the quality factor study has shown advantages of the Light-Hole ground state, we design a further heterostructure where the strengths of both HH and LH ground states are utilized, exploiting the opportunities offered by lattice-matching tensile $\varepsilon$-SiGe barriers to a Ge substrate. 
Building on the demonstrated architecture of $\varepsilon$-Ge bilayer~\cite{tidjaniThreeDimensionalArrayQuantum2025} and the concurrent switchable HH-LH proposal~\cite{GyorgySwitchableHHLH}, here we propose to stack a tensile $\varepsilon$-SiGe quantum well on top of a Ge channel~\cite{costaBuriedUnstrainedGermanium2025} with a $\varepsilon$-SiGe barrier in-between. Specifically, the bottom $18$ nm Ge well is cladded by tensile $\varepsilon$-Si$_{0.12}$Ge$_{0.88}$, with the thin top barrier also acting as intra-well spacer. The top $15$ nm quantum well is tensile $\varepsilon$-Si$_{0.05}$Ge$_{0.95}$, with a top $\varepsilon$-Si$_{0.3}$Ge$_{0.7}$ barrier separating the bilayer from the surface. As shown in Fig.~\ref{fig:deff} c-d), this bilayer stack allows switching the ground state composition between HH and LH by simply changing the applied voltage on the top plunger gate, modeled here via a linear out-of-plane electric field. Fig.~\ref{fig:deff}e) shows the computed linear SOC $\alpha$ as a function of the vertical electric field and displays a three orders of magnitude increase exactly where the ground state switches from HH to LH character. Similarly in Fig.~\ref{fig:deff}f) we extract the planar and out-of-plane g-factor and see the same discontinuous jump from typical HH behaviour with $g_\perp^*>g^*_\parallel$ to the LH inverted setup. In practice, such a heterostructure allows us to exploit the fast operation capabilities of the LH layer while still being able to access the HH ground state.

\textit{Discussion on Experimental Feasibility---} Our theoretical estimates (see details in the SM~\cite{supplemental_material}) of the relaxation of epitaxially grown Ge/SiGe layers shows that our proposed single layer heterostructure sits just above the maximal thickness achievable before relaxation. We note that this limitation can be circumvented by growing a thinner lower barrier along with a larger Ge concentration such as $\text{Si}_{0.12}\text{Ge}_{0.88}$. In the SM~\cite{supplemental_material} we show that such a heterostructure sit well below the critical thickness and yields similar properties. Our proposed bilayer heterostructure is achievable within currently known fabrication constraints and sits below the critical thickness too. 
While we note that for a well with $x_w\gtrsim0.96$ fluctuations in the silicon concentration do not revert the character of the ground state to being HH-dominated, provided that wider wells are implemented, further experimental studies are required to evaluate the impact of the small silicon concentration in the well on the disorder properties of the confined channel. Finally, we stress that such a platform would be lattice matched and hence free from underlying misfit dislocations causing strain fluctuation and cross-hatch patterns in $\varepsilon$-Ge quantum wells for HH qubits\cite{corley-wiciakNanoscaleMapping3D2023}.

\textit{Conclusion}--- We have shown the feasibility and characteristics of a  Light-Hole (LH) ground state in a material stack fully based on Ge/SiGe. We estimate large SOC parameters both linear and cubic in momentum, yielding dipole moments three orders of magnitude larger than $\varepsilon$-Ge Heavy-Hole (HH) ground states. We also show that an isotropic g-tensor can be realized by tuning the design parameters. 
We show that the large gains in dipole moment outweigh the coherence loss and allow giving rise to one order of magnitude increase in quality factor. Finally, we have shown that an extended heterostructure allows for an electrically tunable HH and LH ground state.

\begin{acknowledgments}
\textit{Acknowledgements---}We thank all members of the Bosco, Rimbach-Russ, Scappucci, Veldhorst, and Vandersypen group for valuable feedback. We also thank Zoltán György and Daniel Loss for useful discussions and feedback. This research was sponsored by the Army Research Office under Award Number: W911NF-23-1-0110 and by the EU through H2024 QLSI2.  The views and conclusions contained in this document are those of the authors and should not be interpreted as representing the official policies, either expressed or implied, of the Army Research Office or the U.S. Government. The U.S. Government is authorized to reproduce and distribute reprints for Government purposes notwithstanding any copyright notation herein.
M.R.-R. and E.V. additionally acknowledge support from the Dutch Research Council (NWO) under Award Number Vidi TTW 22204.

\end{acknowledgments}

\clearpage
\begin{widetext}
\section{Supplemental Material for Light-Hole Spin Qubits in Strained SiGe Lattice-Matched to Ge}

In this Supplemential material, we  provide explicit expressions of the necessary elements of the theoretical model used in the main text. We also give additional details on the detailed physical mechanisms that lead to the results in the main text.

\appendix
\section{Model}\label{app:Model description}
We closely follow the methodology in~\cite{vecchioTailoringGermaniumHeterostructures2026} to extract the quantum well and quantum dot properties. We report here a broad outline of the model but refer the reader to references above for a full description of the model. In particular, after diagonalizing the out-of-plane component ($z$-direction) at $B = 0$ and $\textbf{K}_\parallel= 0$ , projecting the full Hamiltonian on the resulting basis gives
\begin{equation}\label{Hkp}
\begin{aligned}
\mathbf{H}_{k\,p}(\mathbf{K}_\parallel) = {}& \mathbf{E}_0^{\mathrm{qw}} + \alpha_0 \left[ \mathbf{M}_1 K_\parallel^2 + \frac{\cos\theta}{2l_B^2}\mathbf{M}_\mathbf{g} + \frac{\sin\theta}{2l_B^2}\left( \frac{\sin\theta}{2l_B^2}\mathbf{N}_\gamma r_\phi - \tilde{\mathbf{N}}_\gamma \right) r_\phi \right] \\
& + \alpha_0 \left[ i\mathbf{M}_1 K_- + \mathbf{M}_2 K_-^2 + \frac{\sin\theta}{2l_B^2}\left( e^{-i\phi}\mathbf{N}_\mathbf{g} + i\mathbf{N}_1^+ r_\phi K_- + i\mathbf{N}_1^- K_- r_\phi \right) + \mathrm{h.c.} \right] \\
& + V_\parallel(x,y),
\end{aligned}
\end{equation}
where $\textbf{K}=k+e/\hbar \textbf{A}$,$K_\pm = K_x \pm i K_y$, $K_\parallel^2 = K_x^2 + K_y^2$, $\{A,B\} = AB+BA$, $r_\phi = x \sin{\phi} - y \cos{\phi}$ and $l_B=\sqrt{\hbar/(eB)}$. The M and N matrices are subband matrices with dimension $n_H\times2 + n_\eta \times2$ with $n_H,n_\eta$ the number of computed Heavy-Hole and Light-Hole/Split-Off subbands respectively, which in typical calculations total to 250 subbands. In particular these states forming the basis can be written as
\begin{equation}
\begin{aligned}
    {}&\ket{Hl;\sigma} = \ket{3/2, 3\sigma/2} \ket{f_l^h}\\
    &\ket{\eta j;\sigma} = \ket{3/2, \sigma/2} \ket{f_j^l}+ \sigma \ket{1/2, \sigma/2} \ket{f_j^s}
    \end{aligned}
\end{equation}
with $l,j$ indices iterating over the respective subbands. This means that for example the $\textbf{E}_0^\text{qw}$ matrix, diagonal by construction in this basis can be written as
\begin{equation}
\textbf{E}_0^\text{qw}=\begin{bmatrix}
    \textbf{E}_0^\text{H} \quad 0 \quad 0 \quad 0 \\
    0 \quad \textbf{E}_0^\eta \quad 0 \quad 0\\
    0 \quad 0 \quad \textbf{E}_0^\eta \quad 0\\
    0 \quad 0 \quad 0 \quad \textbf{E}_0^\text{H}
\end{bmatrix}
\end{equation}
while in general the other matrices can present non-diagonal elements representing the interband coupling which is particularly relevant for the dynamics of the confined holes at the top of the valence band. We refer the reader to~\cite{vecchioTailoringGermaniumHeterostructures2026} for a complete definition of all the matrices above.
\\
The Rashba coefficients can be extracted by projecting Eq.~\eqref{Hkp} onto the LH ground state doublet using a third order Schrieffer-Wolff-Transformation (SWT), and then collecting all the terms that are up to cubic order in k in the resulting effective Hamiltonian. This SWT procedure also provides a quadratic-in-k term which represents the in-plane momentum $\gamma k_\parallel^2$, although the latter only requires second order perturbation to be described exactly~\cite{vecchioTailoringGermaniumHeterostructures2026}. By projecting on the lowest doublet, with $k$ as the perturbation parameter kept up to third order, one obtains the effective Hamiltonian\cite{vecchioTailoringGermaniumHeterostructures2026}
\begin{equation}
    \mathcal{H} = \alpha_0 \gamma k^2_\parallel +  [i(\alpha k_- - \beta_2 k^3_+ - \beta_3 k_- k_+ k_- )\sigma_- + h.c.]
\end{equation}
\\
Finally the g-tensor can be extracted by diagonalizing Eq.~\eqref{Hkp} at $B=0$ with a parabolic confinement in $x$ and $y$ and harmonic basis states ($n_x$=$n_y= 16$), and then projecting the full Hamiltonian on this magnetic field independent eigenbasis
\begin{equation}
\begin{aligned}
\mathbf{\tilde{H}^\text{kp}} = \mathbf{E}_0^{\mathrm{QD}} + \frac{\alpha_0}{2l_B^2}\Bigg\{ & \cos\theta\,\mathbf{L}_{2\perp} + \frac{\cos^2\theta}{2l_B^2}\mathbf{L}_{4\perp} + \frac{\sin^2\theta}{2l_B^2}\mathbf{L}_{4\|} + \sin\theta\left[ e^{-i\phi}\mathbf{L}_{2\|} + \frac{\sin\theta}{2l_B^2}e^{-2i\phi}\mathbf{L}'_{4\|} + \frac{\cos\theta}{2l_B^2}e^{-i\phi}\mathbf{L}_{4\times} + \mathrm{h.c.} \right] \Bigg\},
\end{aligned}
\end{equation}
where $\mathbf{E}_0^{\mathrm{QD}}$ are the eigenvalues of the quantum dot (QD) at $B=0$ from the previous step and the $\mathbf{L}$-matrices are

\begin{align}\label{L matrices}
\mathbf{L}_{2\perp} &= \mathbf{V}^\dagger\left[\mathbf{M}_g + 2\mathbf{M}_\gamma L_z + (2\mathbf{M}_1\rho_- - 4i\mathbf{M}_2\rho_- k_- + \mathrm{h.c.})\right]\mathbf{V}, \\
\mathbf{L}_{2\|} &= \mathbf{V}^\dagger\left[\mathbf{N}_g - i\mathbf{N}'_\gamma\rho_+ - \mathbf{N}_1\{\rho_+, k_-\} + 2\mathbf{N}_1^\dagger\rho_+ k_+\right]\mathbf{V}, \\
\mathbf{L}_{4\perp} &= \mathbf{V}^\dagger\left[4\mathbf{M}_\gamma\rho_-\rho_+ - 4\left(\mathbf{M}_2\rho_-^2 + \mathrm{h.c.}\right)\right]\mathbf{V}, \\
\mathbf{L}_{4\|} &= \mathbf{V}^\dagger\left[2\mathbf{N}_\gamma\rho_-\rho_+\right]\mathbf{V}, \\
\mathbf{L}'_{4\|} &= \mathbf{V}^\dagger\left[-\mathbf{N}_\gamma\rho_+^2\right]\mathbf{V}, \\
\mathbf{L}_{4\times} &= \mathbf{V}^\dagger\left[4i\left(\mathbf{N}_1\rho_-\rho_+ + \mathbf{N}_1^\dagger\rho_+^2\right)\right]\mathbf{V},
\end{align}
where $L_z = xk_y - yk_x$ and $\mathbf{V}$ is the matrix of (column) eigenvectors of the QD at $B=0$. Ultimately the g-tensor components can be read out as the terms of the L matrices linear in magnetic field.

\section{Subband Parameters}\label{app:subband}
In this section we further discuss the intuition behind the mechanisms that drive the behaviour of the subband parameters in the main text. In particular we focus on the linear term $\alpha$ and $\beta_2$. For the former, under the assumption of an infinite well and a 6-band Luttigner-Kohn model for the valence band, an analytic expression can be derived~\cite{delvecchioDynamicsSpinOrbitCoupling2024}
\begin{equation}
    \alpha = -6 \sqrt{2} \alpha_0 \gamma_3 \mathcal{I}\{ \bra{f^l}k_z \ket{f^s}\}
\end{equation}
with $\alpha_0 =\frac{\hbar^2}{2 m_0}$ a constant, $\gamma_3$ the Luttinger parameter, $k_z$ the momentum operator in the growth direction. Even though this expression strictly holds only for an infinite well, we can plot the main constituents for the heterostructure discussed in the main text to provide a physical intuition of the observed trends.
\begin{figure*}
    \centering
    \includegraphics[width=0.45\linewidth]{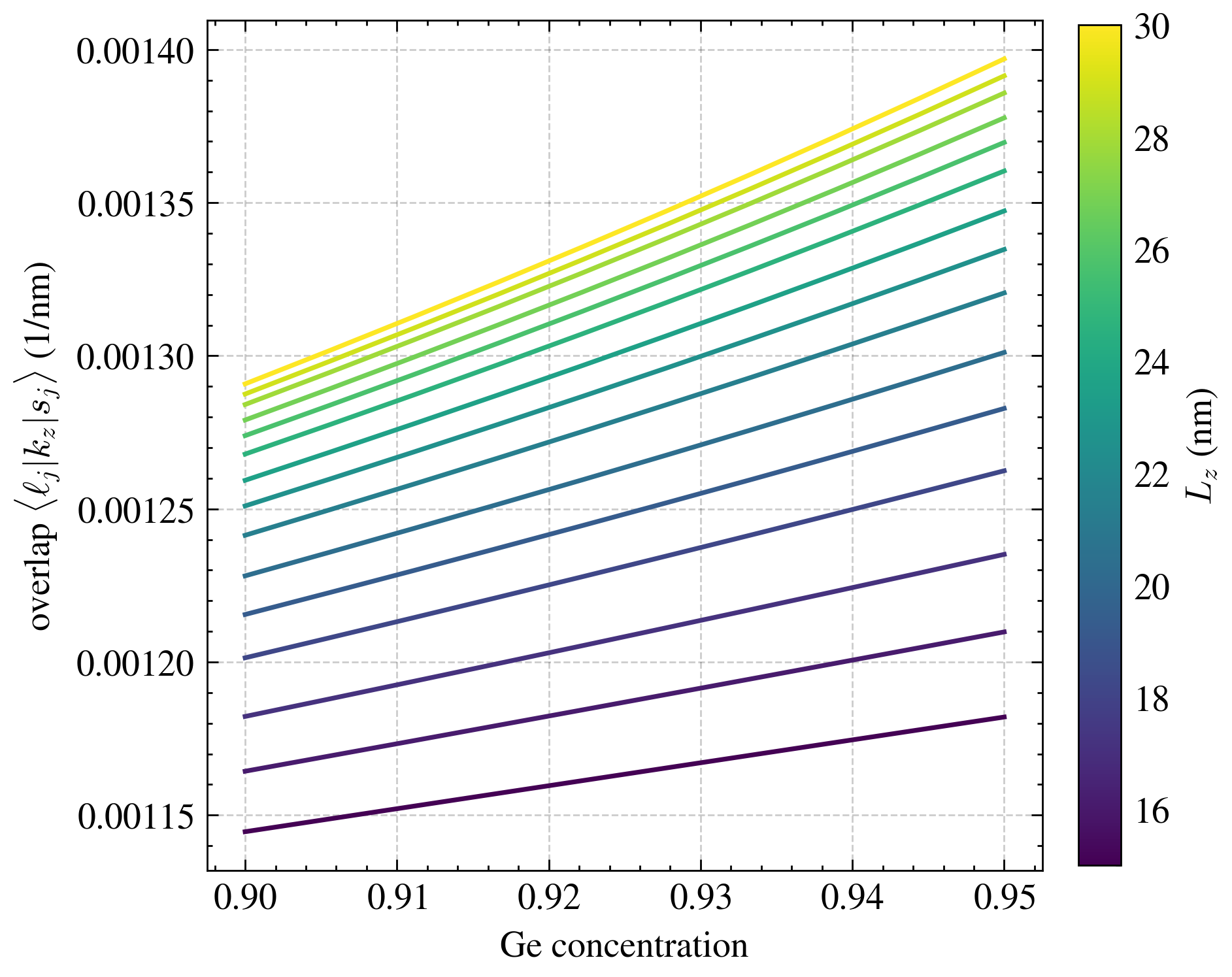}
    \includegraphics[width=0.45\linewidth]{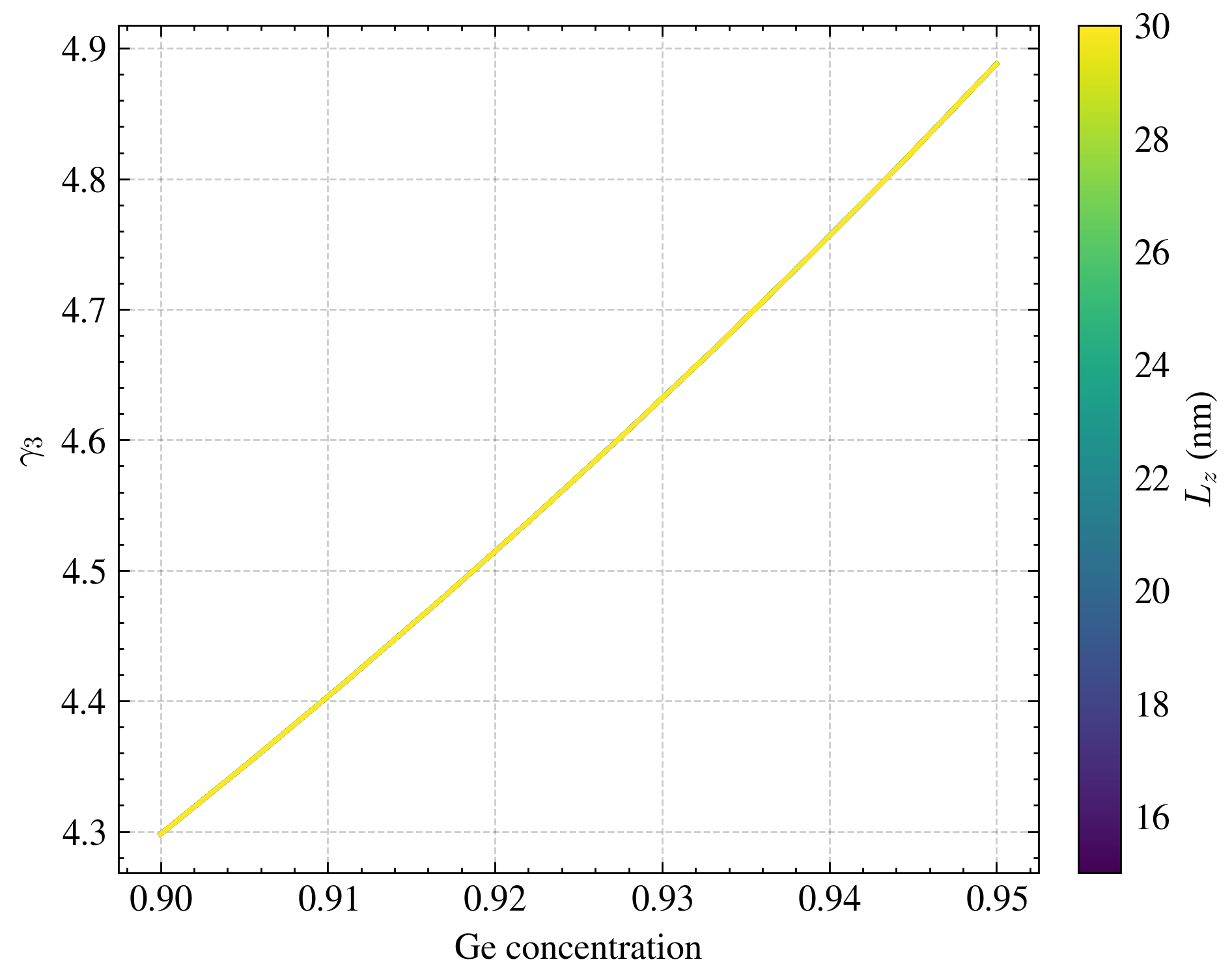}
    \caption{(a) Light-Hole/Split-Off overlap as a function of the Ge concentration in the well and well thickness. (b) Luttinger parameter $\gamma_3$. We plot these quantities as a function of Ge concentration in the well and well thickness. These quantities were extracted for a vertical field of 1 MV/m.}
    \label{fig:alpha_constituents}
\end{figure*}
Fig.~\ref{fig:alpha_constituents} reveals that the LH/SO overlap is strongly dependent on the thickness of the well and an increase in Ge concentration also boosts the Luttinger parameter $\gamma_3$.
For the cubic Rashba $\beta_2$ we can again follow the intuition provided by the approximate expressions for an infinite QW. In particular, keeping only the second order terms in the perturbative expansion, and neglecting the split-off band contributions (effectively setting the $f^s =0$)~\cite{delvecchioDynamicsSpinOrbitCoupling2024}
\begin{equation}
    \beta_2 =3 i \alpha_0^2 (\gamma_2 - \gamma_3)\gamma_3 \bra{f^l_0}[k_z,\sum_l\frac{\ket{f^H_l}\bra{f^H_l}}{E_0 - E_l}\ket{f^l_0}]
\end{equation}
where $E_0$ denotes the energy of the lowest Light-Hole subband and $f^H$ is the label of the Heavy-Hole subbands.
\begin{figure*}
    \centering
    \includegraphics[width=0.5\linewidth]{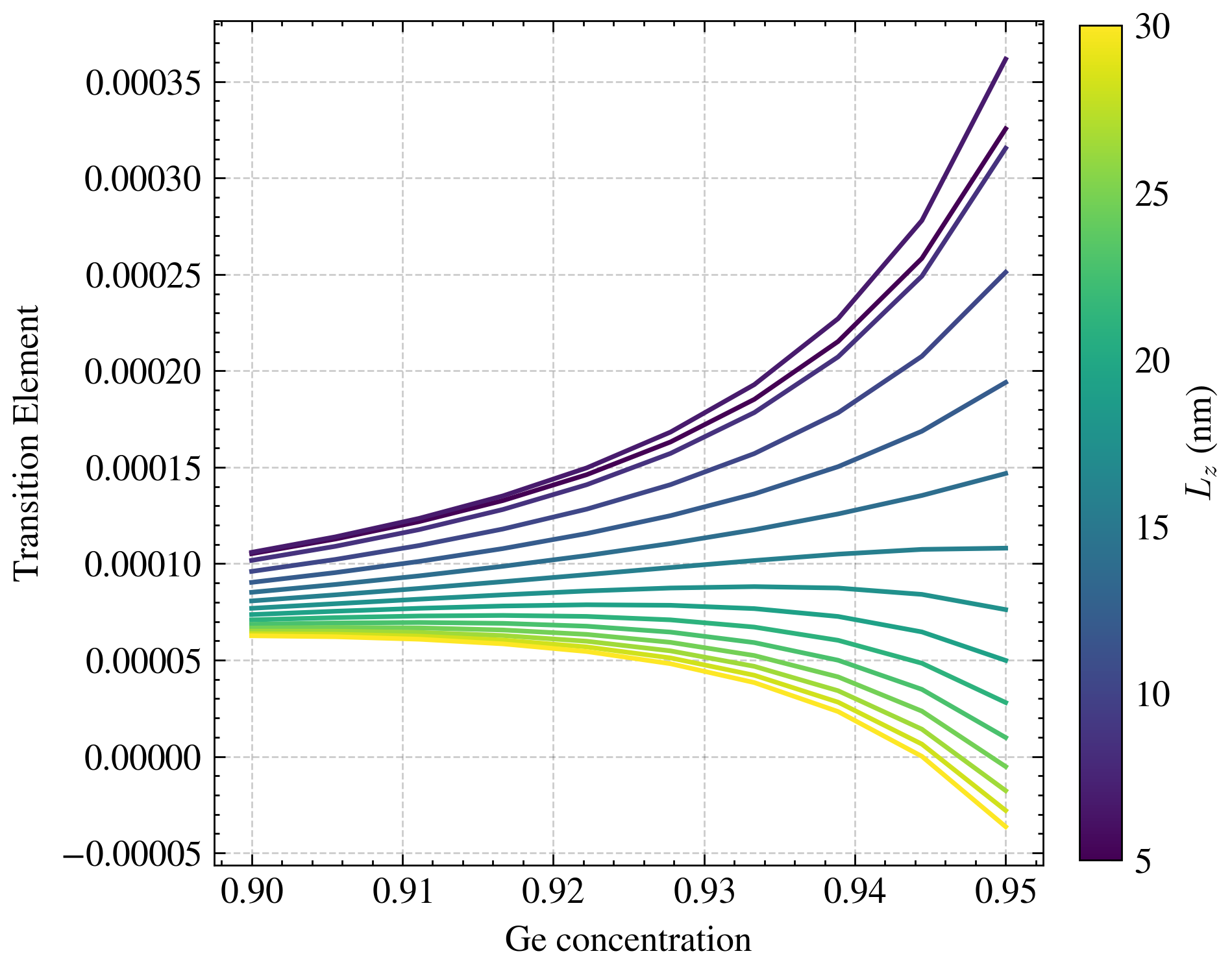}
    \caption{Transition matrix element $\bra{f^L_0}[k_z,\sum_l\frac{\ket{f^h_l}\bra{f^h_l}}{E_0 - E_l}\ket{f^L_0}$ plotted as a function of the Ge concentration in the well and well thickness. This quantity was extracted for a vertical field of 1 MV/m. }
    \label{fig:transition_matrix_element_beta2}
\end{figure*}
In Fig.~\ref{fig:transition_matrix_element_beta2} we observe the behavior of this expression and the resemblance with the trends observed in the main text. We note however that for both $\beta_2$ and $\beta_3$ the third order Schrieffer-Wolff terms are especially relevant and dominate the dynamics of $\beta_3$ with respect to the first and second order terms~\cite{vecchioTailoringGermaniumHeterostructures2026}.

\section{g-Tensor components}\label{app:g}
We can intuitively understand the non-monotonic behaviour of the out of plane g-factor of the main text by considering the HH-LH coupling in the approximate formula for the correction of the out of plane g-factor in the case of an infinite QW without planar confinement~\cite{delvecchioDynamicsSpinOrbitCoupling2024}:
\begin{equation}
     g_\perp^* \propto 2\alpha_0\left(\sum_l \frac{T_{l,j}^{\times*}T_{l,j}^{\times}}{E_j^{\eta} - E_l^{\mathrm{H}}}\right)
\end{equation}
with $T^\times$ the HH-LH coupling subband matrix~\cite{vecchioTailoringGermaniumHeterostructures2026}. In particular Fig.~\ref{fig:Tx} shows a similar non-monotonic trend as observed in the main text.

\begin{figure}
    \centering
    \includegraphics[width=0.5\linewidth]{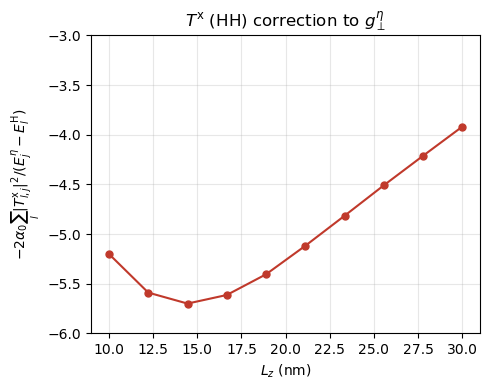}
    \caption{Plot of the perturbative correction to the out-of-plane g-factor for a LH ground state under the infinite QW approximation for a sweep of well thicknesses}
    \label{fig:Tx}
\end{figure}
To understand the trend and strong tunability of the out-of-plane g-factor with respect to the electric field we decompose each term of the computed g-factor.
\begin{figure}
    \centering
    \includegraphics[width=0.5\linewidth]{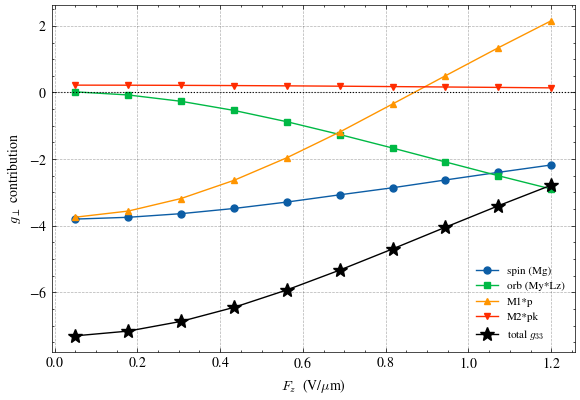}
    \caption{Decomposition of the out-of-plane g factor computed as Eq.~\eqref{L matrices} (terms composing $\mathbf{L}_{2\perp}$) as a function of out-of-plane electric field.}
    \label{fig:gperp decomposition}
\end{figure}
From Fig~\ref{fig:gperp decomposition} we see that the HH-LH coupling matrix elements in $M_1 \rho_-$ are one of the dominant terms for the strong renormalization of the out-of-plane g-factor. Moreover in \ref{fig:T_e} we observe the increase in $J_z = \pm1/2$ which mixes the $B=0$ ground state composition. This change in the eigenstate composition changes the Zeeman contribution directly.
\begin{figure}
    \centering
    \includegraphics[width=0.8\linewidth]{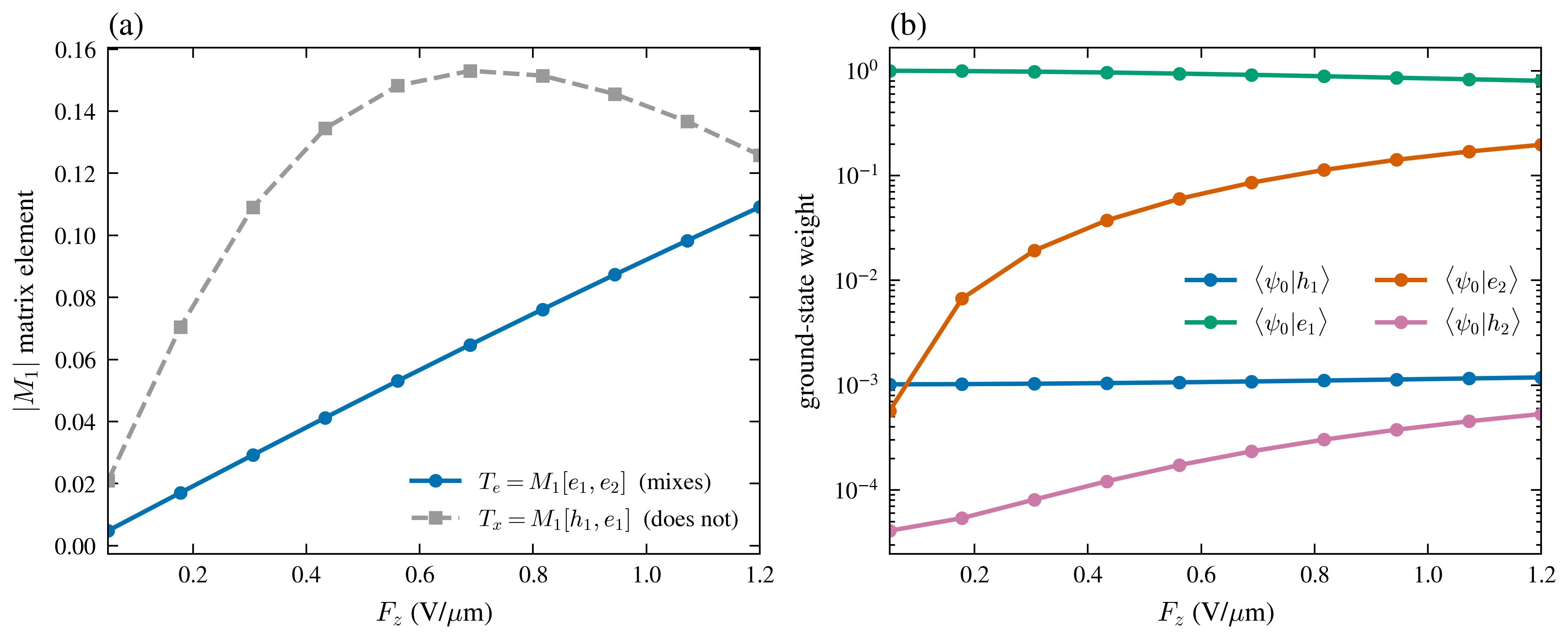}
    \caption{(a) Decomposition of the largest $T^\times$ and $T^\eta$ element of $M_1$ that give rise to a strong mixing of opposite momentum terms in the ground state.(b) ground state composition with respect to the $J_z$ eigenstates with the labels $h_{1,2}$ and $e_{1,2}$ corresponding to the $\pm$ HH and $\eta$ subbands respectively.}
    \label{fig:T_e}
\end{figure}
At high field the ground state stops being an eigenstate of $J_z$ because the intraband LH/SO coupling is increased and the ground state, via linear spin orbit coupling, becomes a superposition of $\ket{\eta_1}$ and $\ket{\eta_2}$.

\section{Heterostructure with relaxed growth constraints}
We have highlighted in the main text that the proposed single layer structure sits beyond the critical thickness regime the grown heterostructure can start to relaxe. However, a Light-Hole ground state is still achieved by lowering the Si concentration of the bottom SiGe barrier layer. To quantify the allowed regime we estimate the critical thickness of the grown heterostructure before relaxation, we approximately estimate the critical thickness from the energy-balance criterion of
People and Bean~\cite{peopleCalculationCriticalLayer1985} and generalise it to a multilayer heterostructure. In the calculations we further assume dislocations to take place at the interface of the substrate with the first barrier layer. The self-energy is consequentally
set by the total stack thickness $L=\sum_i h_i$, while the stored strain
energy sums over the individual layers
\begin{equation}
  \sum_i f_i^{2} h_i = C \ln\!\left(\frac{L}{b}\right),
  \qquad
  C = \left(\frac{1-\nu}{1+\nu}\right)
      \frac{b^{2}}{16\pi\sqrt{2}\,\langle a\rangle}.
  \label{eq:criterion}
\end{equation}
Here $f_i = \left[a_\mathrm{sub}-a(x_i)\right]/a(x_i)$ is the misfit of layer
$i$, $b=a_\mathrm{sub}/\sqrt{2}$ the slip distance, $\langle a\rangle$ the
thickness-weighted mean lattice constant, and $\nu=0.28$. The relaxed lattice
constants follow $a(x)=5.4307+0.1992\,x+0.02733\,x^{2}$~\AA~\cite{dismukesLatticeParameterDensity1964} and for a relaxed Ge substrate, $b=4.00$~\AA{} and $C=2.24\times10^{-2}$~\AA. The
structure does not relax as long as the left-hand side of Eq.~\eqref{eq:criterion} remains below the right.

The curve in Fig.~\ref{fig:hc} gives the maximum
bottom-barrier thickness within the full
Si$_{1-x}$Ge$_{x}$/Si$_{0.05}$Ge$_{0.95}$/%
Si$_{0.30}$Ge$_{0.70}$ stack, obtained by solving
Eq.~\eqref{eq:criterion} using bisection with respect to thickness of the bottom layer keeping the remaining layers fixed. The top barrier thickness is set to $50$ nm ensuring that the well is buried from the surface and defects at the oxide interface. The well thickness is kept at $25$ nm, the maximum well thickness considered in the main text study. We have added three markers. The first (cross) shows the single layer heterostructure proposed in the main text, and the second (circle) for a design where the bottom spacer is only $30$ nm thick and is Si$_{0.12}$Ge$_{0.88}$, sitting comfortably below the critical thickness. The third marker (square) corresponds to the bilayer proposed the main text and sits below; the bottom Ge layer does not contribute to relaxation since it is lattice matched, the two Si$_{0.12}$Ge$_{0.88}$ barriers have a total thickness of $30$ nm, and the tensile strained well is only $15$ nm thick.

In Fig.~\ref{fig:bottomSiGe88} we plot the properties of the new heterostructure with thinner bottom layer with higher Ge concentration along with the computed HH-LH gap. We see quantitatively similar results compared to the results in the main text; a large spin-orbit interaction and a resilient LH ground state. In particular, in Fig.~\ref{fig:Rashba bottomSiGe88} we compute the Rashba parameters and see a quantitative similar response compared to the results of the main text except for $\beta_2$. However, we remark that $\beta_2$ is much smaller and less relevant. Interestingly, the effect of the well thickness is less relevant compared to the heterostructure shown in the main text as expected for a lower bottom barrier height.
\begin{figure}
    \centering
    \includegraphics[width=0.8\linewidth]{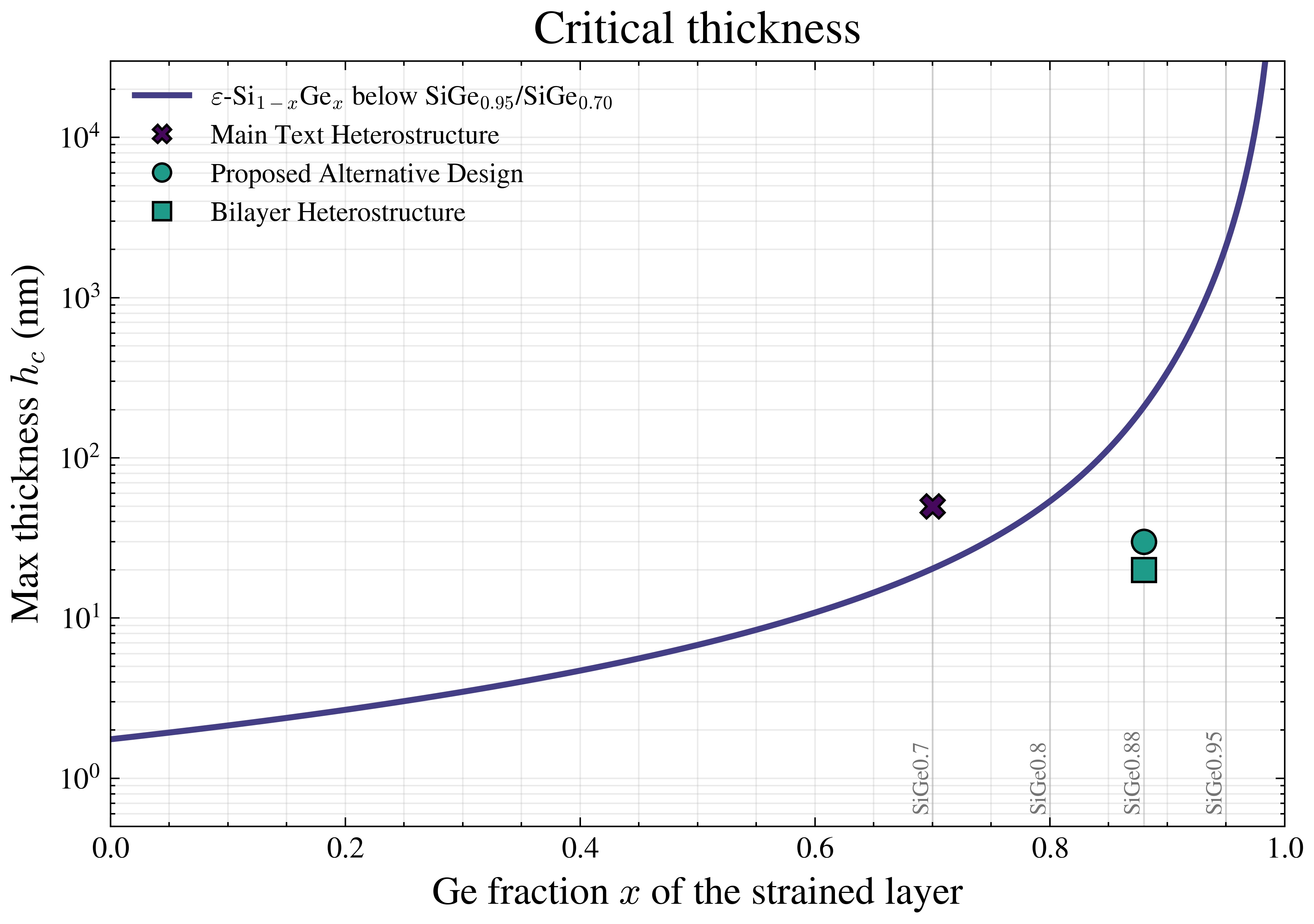}
    \caption{Maximum bottom-barrier thickness for the full Si$_{1-x}$Ge$_{x}$/Si$_{0.05}$Ge$_{0.95}$/Si$_{0.30}$Ge$_{0.70}$ stack. The markers display the position of the proposed architectures, single layer structure, alternative single layer structure and proposed bilayer, calculated from the thickness of the bottom Si$_{0.30}$Ge$_{0.70}$, Si$_{0.12}$Ge$_{0.88}$ and Si$_{0.12}$Ge$_{0.88}$ barriers.}
    \label{fig:hc}
\end{figure}

\begin{figure}
    \centering
    \includegraphics[width=0.8\linewidth]{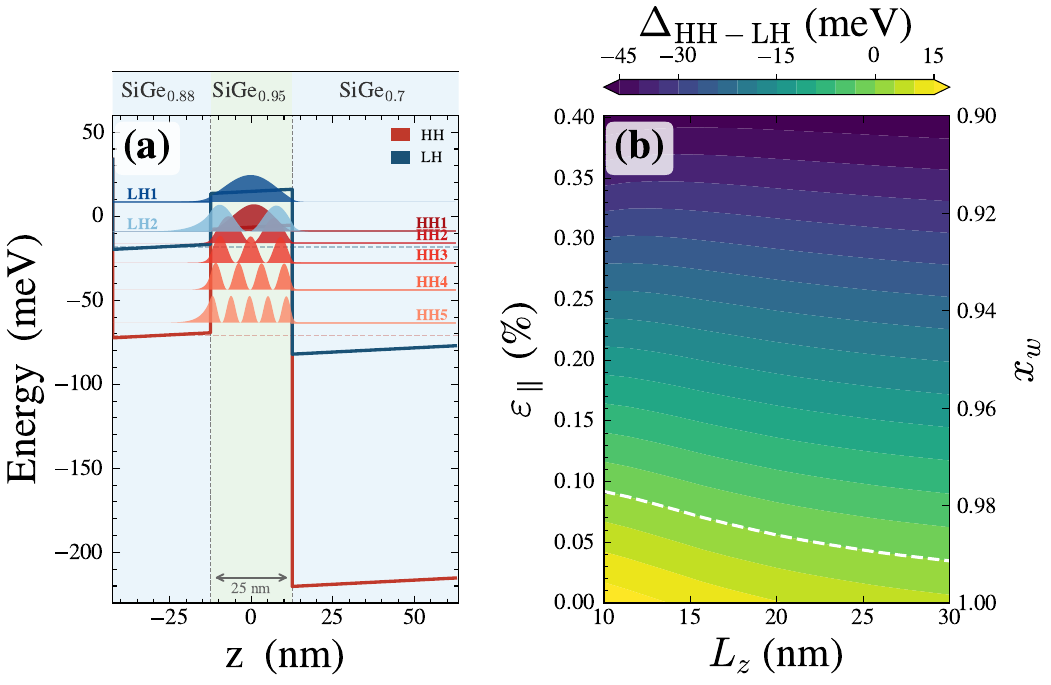}
    \caption{(a) Schematics of the alternative Heterostructure and (b) HH-LH gap for a system with thinner bottom barrier with increased Ge concentration to reduce the risk of relaxation during the growth}
    \label{fig:bottomSiGe88}
\end{figure}

\begin{figure}
    \centering
    \includegraphics[width=0.8\linewidth]{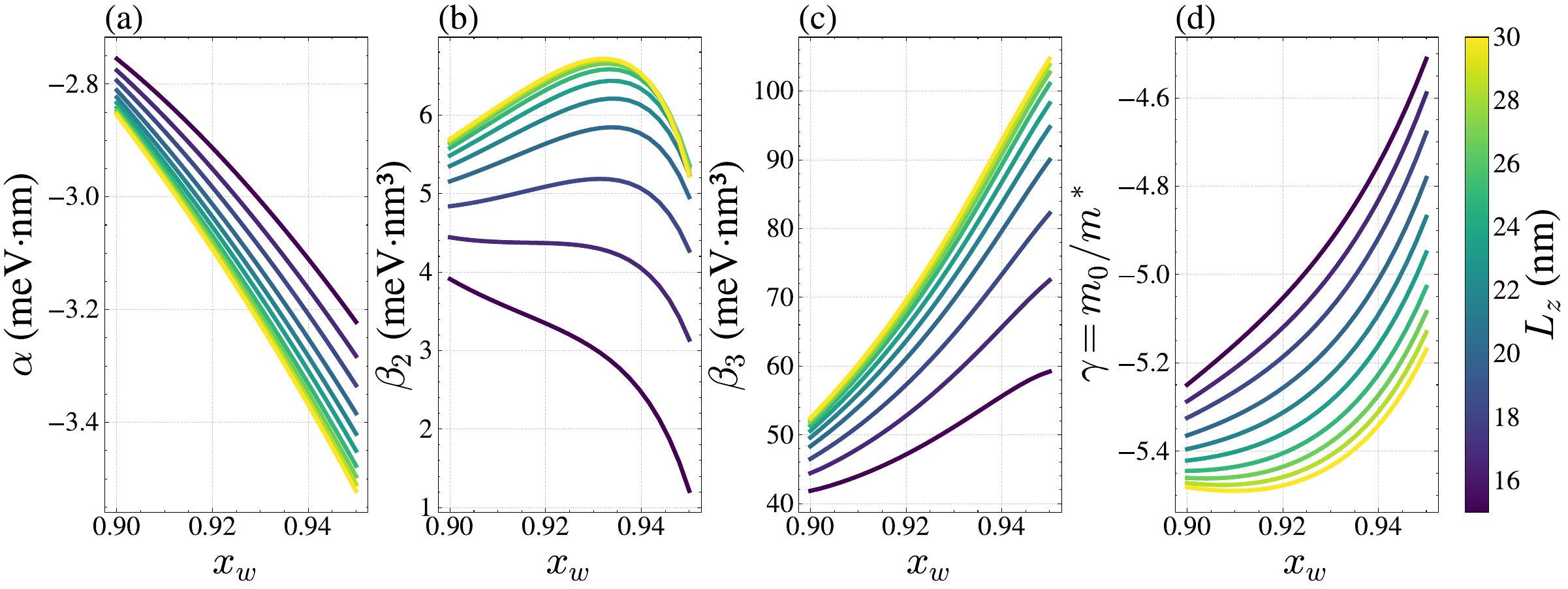}
    \caption{Rashba coefficients and effective mass of the modified heterostructure as a function of well thickness and Ge well concentration $x_w$. The parameters were extracted for an out-of-plane electric field of $1$ MV/m.}
    \label{fig:Rashba bottomSiGe88}
\end{figure}
	
\end{widetext}

\bibliography{apssamp}
\end{document}